\documentclass[a4paper, journal]{IEEEtran}
\usepackage[cmex10]{amsmath}
\usepackage{hhline}
\usepackage{amsmath,graphicx,multirow}
\usepackage{times,epsfig}
\usepackage{bm,amssymb}
\usepackage{color, soul}
\usepackage{algorithm,algpseudocode}
\usepackage{eqparbox}
\usepackage{epstopdf}
\usepackage{framed}
\usepackage{array}
\usepackage{mdwmath}
\usepackage{mdwtab}
\usepackage{amscd}
\usepackage[tight,footnotesize]{subfigure}

\begin{document}
%
\title{\huge{Fast Signal Separation of 2D Sparse Mixture \\via Approximate Message-Passing}}

\author{ Jaewook~Kang,~\IEEEmembership{Student  Member,~IEEE,}
        Hyoyoung~Jung,~\IEEEmembership{Student Member,~IEEE,}\\
        and~Kiseon~Kim,~\IEEEmembership{Senior Member,~IEEE}


 \vspace{-20pt}

\thanks{Manuscript received May 6, 2015; revised July 2, 2015; accepted July 5, 2015. The associate editor coordinating the review of this manuscript
and approving it for publication was Dr. Jun Fang. This research was
a part of the project titled 'Development of Ocean Acoustic Echo
Sounders and Hydro-Physical Properties Monitoring Systems', and
'Development of an Integrated Control System of Eel Farms based on
Short-range Wireless Communications' funded by the Ministry of
Oceans and Fisheries, Korea. }
  \thanks{Copyright (c) 2015 IEEE. Personal use of this material is permitted. However, permission to use this material
  for any other purposes must be obtained from the IEEE by sending a request to pubs-permissions@ieee.org.}
\thanks{The authors are with Department of Information and Communication,
Gwangju Institute of Science and
 Technology, Gwangju, Republic of Korea
 (Email:\{jwkkang,rain,kskim\}@gist.ac.kr)}
}
\markboth{To Appear in IEEE Signal Processing Letters,  VOL. 22,
Issue 11, Nov. 2015}{J.Kang \MakeLowercase{\textit{et al.}}}
 \maketitle

\begin{abstract}
Approximate message-passing (AMP) method is a simple and efficient
framework for the linear inverse problems. In this letter, we
propose a faster AMP to solve the \emph{$L_1$-Split-Analysis} for
the 2D sparsity separation, which is referred to as \emph{MixAMP}.
We develop the MixAMP  based on the factor graphical modeling and
the min-sum message-passing. Then, we examine MixAMP for two types
of the sparsity separation: separation of the direct-and-group
sparsity, and that of the direct-and-finite-difference sparsity.
This case study shows that the MixAMP method offers computational
advantages over the conventional first-order method, TFOCS.
\end{abstract}


\section{Introduction}
The linear inverse problems for estimating an unknown  signal
$\mathbf{X}$ from linear measurements $\mathbf{Y}$  are central in
signal processing techniques. In this letter, we revisit such
inverse problems with three additives:
\begin{itemize}
\item  $\mathbf{X}$ is a two-dimensional (2D) square, whose
entries are with the index pair $(i,j) \in \Lambda :=
\{1,...,\sqrt{N}\}\times\{1,...,\sqrt{N}\}$ where $\sqrt{N} \in
\mathbb{Z}^+$ and $N:=|\Lambda|$,
\item $\mathbf{X}$  is a mixture of  two  signals $\mathbf{X}_a, \mathbf{X}_b$ having sparse
representation in two separate basis, \emph{i.e.},
$\mathbf{X}=\mathbf{X}_a+\mathbf{X}_b$.
\item $\mathbf{Y}$  are \emph{Compressed Sensing} (CS)  measurements such that $\mathbf{Y}$ includes its effective samples with  the index pair $(k,l) \in \Omega\subseteq
\Lambda$ where $M:=|\Omega|$.
\end{itemize}
From the 2D setup, we generate the linear measurements  by
\begin{align}\label{lin_sys}
\mathbf{Y}=\mathcal{P}_{\Omega}\left\{\mathbf{A} \mathbf{X}
\mathbf{A}^T \right\} \in \mathbb{R}^{\sqrt{N} \times \sqrt{N}},
\end{align}
where $\mathbf{A} \in \mathbb{R}^{\sqrt{N} \times \sqrt{N}}$ is a
 measurement matrix; $\mathcal{P}_{\Omega}\{\cdot\}$ is the
undersampling operator nulling entries of its matrix argument not in
the set $\Omega$. The  measurement model \eqref{lin_sys} is
motivated by the 2D signal acquisition concept considering two
perpendicular spatial/spectral axes of $\mathbf{X}$ where ``the left
$\mathbf{A}$" linearly mixes  $\mathbf{X}$ in vertical  axis and
``the right $\mathbf{A}^T$" does in  horizontal axis. As practical
applications of \eqref{lin_sys}, the 2D Fourier-transform-based
image/video acquisition \cite{bregman},\cite{videoAMP} and optical
image encryption \cite{2D-CS} have been considered.

The measurement model \eqref{lin_sys} has two advantages in 2D
signal processing. This 2D model spends $\mathcal{O}(N^{3/2})$
complexity for the measurement generation, which is less  than
$\mathcal{O}(N^2)$ by the 1D equivalent model \cite{Marco},
\begin{align}\label{lin_sys_vec}
{\text{vec}}({\mathbf{Y}}) =
\mathcal{P}_{\Omega'}\left\{({\mathbf{A}} \otimes
{\mathbf{A}})\,{\text{vec}}( \mathbf{X})\right\} \in
{\mathbb{R}^{{N} \times 1}},
\end{align} where $\otimes$ is Kronecker product, $\text{vec}(\cdot)$
is the columnwise vectorizing operator, and
$\Omega'\subseteq\{1,...,N\}$. In addition, the 2D  model is storage
efficient by saving memory for the Kronecker product matrix
${\mathbf{A}} \otimes {\mathbf{A}} \in \mathbb{R}^{ N \times N}$
which is unavoidable under the 1D model \eqref{lin_sys_vec}.

Our main task is to simultaneously estimate the two 2D signals,
$\mathbf{X}_a,\mathbf{X}_b$,  given the knowledge of $\mathbf{Y}$,
$\mathbf{A}$, and $\Omega$ under \eqref{lin_sys}. This problem
so-called \emph{Sparse signal separation}, being related to the
\emph{Analysis CS} whose unknown signal is sparse in a concatenation
of the two basis \cite{candes}, including applications to image
inpainting and deblurring \cite{deblurring}, and super-resolution
\cite{mallat}.

When the cardinality of the sets has $|\Omega| < |\Lambda|$ such
that the problems are ill-posed, optimization methods with
regularization have been mostly considered. A known approach to the
problem is the \emph{$L_1$-Split-Analysis}
\cite{candes}-\cite{deblurring}:
\begin{align}\label{L1min}
\,\left\{ \begin{gathered}
  \mathop {\min }\limits_{\mathbf{X}_a,\mathbf{X}_b} \,\,||{{\mathcal{T}}_a}\mathbf{X}_a||_1 + ||{{\mathcal{T}}_b}\mathbf{X}_b||_1  \hfill \\
  {\text{s.t. }}  \mathbf{Y}=\mathcal{P}_{\Omega}\left\{\mathbf{A}
(\mathbf{X}_a+\mathbf{X}_b)\mathbf{A}^T \right\},
 \hfill
\end{gathered}  \right.
\end{align}
where the  sparsity of $\mathbf{X}_a,\mathbf{X}_b$ is promoted  in
terms of analysis transform operators
${{\mathcal{T}}_a},{{\mathcal{T}}_b}$. This $L_1$-Split-Analysis is
related to \emph{Morphological Component Analysis} (MCA)
\cite{donoho},\cite{MCA} in that the both analysis approaches
decompose $\mathbf{X}$ by pursuing sparse representation of
$\mathbf{X}_a,\mathbf{X}_b$. However, they are distinct in that MCA
does not include measurement compression such as \eqref{lin_sys} but
simply goes with $\mathbf{Y}=\mathbf{X}_a+\mathbf{X}_b$.

Practical solving of \eqref{L1min} has been considered in the works
of \cite{aubel},\cite{deblurring} via the \emph{Templates for
First-Order Conic Solvers} (TFOCS)  \cite{TFOCS} and the \emph{Split
Bregman iteration} \cite{bregman}, respectively. However, their
examples are not in the context of our problem setup since they do
not contain the concept of linear mixing by setting
$\mathbf{A}=\mathbf{I}$.

In this letter, we propose an \emph{Approximate Message-Passing}
(AMP) method for solving the \emph{$L_1$-Split-Analysis} under the
2D CS model \eqref{lin_sys}. This is motivated by excellent
properties of AMP \cite{AMP},\cite{TVAMP}: \emph{i)}  asymptotic
Lasso performance, \emph{ii)} efficient computations, and
\emph{iii)} algorithmic simplicity. We refer to the proposed method
as \emph{separation of sparse mixture  via approximate
message-passing} (MixAMP). We claim that MixAMP  is remarkably
faster than the conventional first-order method,  TFOCS
\cite{TFOCS}, for the 2D sparse signal separation task.

We believe that another advantage of this MixAMP lies in its
flexibility. The 2D separation problems, regarding various types of
the sparsity, can be solved via the MixAMP once a proper denoiser is
given. In the sequel, we examine the MixAMP  for two cases of the
sparsity separation, \emph{i)} separation of direct-and-group sparse
mixture, and \emph{ii)}  that of direct-and-finite-difference (FD)
sparse mixture, by applying the simple denoisers introduced in the
literature \cite{AMP},\cite{TVAMP}. In each case, we demonstrate the
low-computationality of MixAMP by an exemplary comparison to the
TFOCS method.\footnote{The MATLAB codes for this comparison is
available at our website, https://sites.google.com/site/jwkang10/.}

\begin{figure}
\centering
\includegraphics[width=9cm]{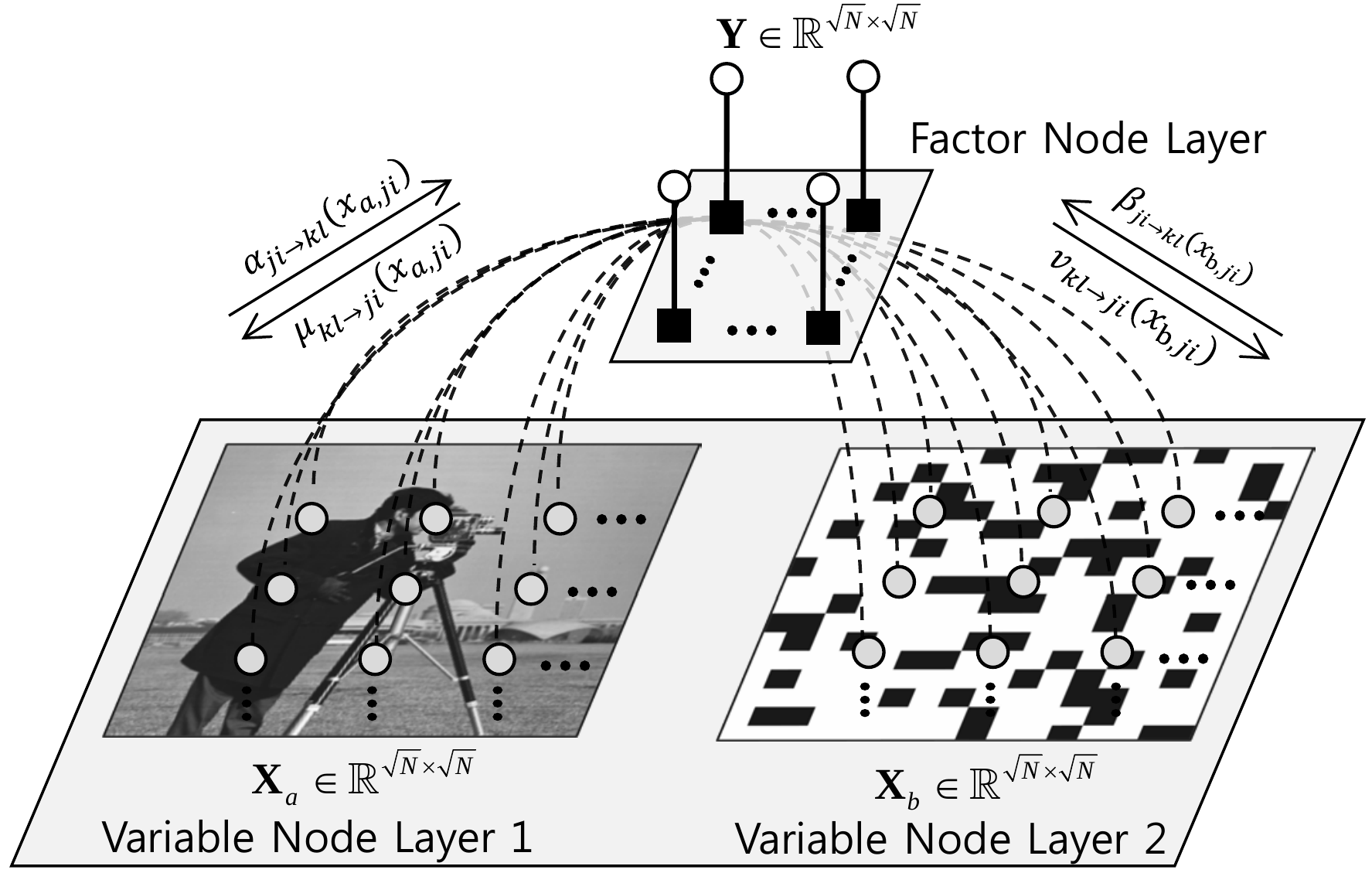}
\caption{Factor graphical model and message-passing  for the
$L_1$-Split-Analysis problem \eqref{L1min}.} \label{fig1}
\end{figure}

\section{AMP Method for $L_1$-Split-Analysis}
The first step of the AMP development is to establish a
message-passing rule based on a factor graphical model of linear
systems. Fig.\ref{fig1} describes a factor graph  for our system
\eqref{lin_sys}, deriving the corresponding joint PDF, given by
\begin{align}\label{jointPDF}
{f_{{{\mathbf{X}}_a},{{\mathbf{X}}_b},{\mathbf{Y}}}}( \cdot) \propto
{f_{{{\mathbf{X}}_a}}}(\cdot;{\mathcal{T}}_a){f_{{{\mathbf{X}}_b}}}(\cdot;{\mathcal{T}}_b)
\prod\limits_{(k,l)\in\Omega}
{{f_{{Y_{kl}}|{{\mathbf{X}}_a},{{\mathbf{X}}_b}}}(\cdot)}
\end{align}
where the two priors,
${f_{{{\mathbf{X}}_a}}}(\cdot;{\mathcal{T}}_a)$ and
${f_{{{\mathbf{X}}_b}}}(\cdot;{\mathcal{T}}_b)$, sparsify the
signals in the domain of ${\mathcal{T}}_a$ and ${\mathcal{T}}_b$,
respectively. From the joint PDF \eqref{jointPDF}, we form a
message-passing rule for solving \eqref{L1min} in the manner of the
\emph{min-sum algorithm} \cite{Montanari}, given by
\begin{align}\label{minsum}
& {\text{a) For the signal }}{{\mathbf{X}}_a}:\forall ( ji,kl) \in   \Lambda   \times \Omega  \nonumber  \\
&\footnotesize
  \left\{ \begin{gathered}
  {\mu_{kl \to ji}}({x_{a,ji}}) = \mathop {\min }\limits_{ \begin{subarray}{l}\{ {\mathbf{x}}_a,\mathbf{x}_b\} \\\backslash {x_{a,ji}}\end{subarray}}  \log {f_{{Y_{kl}}|{{\mathbf{X}}_a},{{\mathbf{X}}_b}}}( \cdot ) + \sum\limits_{\begin{subarray}{l} j'i' \\ \ne ji\end{subarray}} {\alpha_{j'i' \to kl}}({x_{a,j'i'}})  \hfill \\
  \,\,\,\,\,\,\,\,\,\,\,\,\,\,\,\,\,\,\,\,\,\,\,\,\,\,\,\,\,\,\,\,\,\,\,\,\,\,\,\,\,\,\,\,\,\,\,\,\,\,\,\,\,\,\,\,\,\,\,\,\,\,\,\,\,\,\,\,\,\,\,\,\,\,\,\,\,\,\,\,\,\,\,\,\,\,\,\,\,\,\,\,\,\,\,\,\,\,\,\,\,\, \,\,\,\,\,  + \sum\limits_{ji\in\Lambda} {{\beta_{ji \to kl}}({x_{b,ji}})}  \hfill \\
  {\alpha_{ji \to kl}}({x_{a,ji}}) = \log {f_{{{\mathbf{X}}_a}}}( \mathbf{x}_a ;{\mathcal{T}}_a) + \sum\limits_{k'l' \ne kl} {{\mu_{k'l' \to ji}}({x_{a,ji}})}  \hfill \\
\end{gathered}  \right.  \nonumber \\
&{\text{b) For the signal }}{{\mathbf{X}}_b}:\forall  (ji,kl)  \in  \Lambda   \times \Omega  \nonumber\\
&\footnotesize
  \left\{ \begin{gathered}
  {\nu_{kl \to ji}}({x_{b,ji}}) = \mathop {\min }\limits_{ \begin{subarray}{l}\{ {\mathbf{x}}_a,\mathbf{x}_b\} \\\backslash {x_{b,ji}}\end{subarray}} \log {f_{{Y_{kl}}|{{\mathbf{X}}_a},{{\mathbf{X}}_b}}}( \cdot ) + \sum\limits_{ji\in \Lambda} {{\alpha_{ji \to kl}}({x_{b,ji}})}  \hfill \\
  \,\,\,\,\,\,\,\,\,\,\,\,\,\,\,\,\,\,\,\,\,\,\,\,\,\,\,\,\,\,\,\,\,\,\,\,\,\,\,\,\,\,\,\,\,\,\,\,\,\,\,\,\,\,\,\,\,\,\,\,\,\,\,\,\,\,\,\,\,\,\,\,\,\,\,\,\,\,\,\,\,\,\,\,\,\,\,\,\,\,\,\,\,\,\,\,\,\,\,\,\,\, \,\, + \sum\limits_{j'i'\ne ji} {{\beta_{j'i' \to kl}}({x_{b,j'i'}})}  \hfill \\
  {\beta_{ji \to kl}}({x_{b,ji}}) = \log {f_{{{\mathbf{X}}_b}}}(\mathbf{x}_b  ;{\mathcal{T}}_b) + \sum\limits_{
  k'l'\ne kl} {{\nu_{k'l' \to ji}}({x_{b,ji}})},  \hfill \\
\end{gathered}  \right.
\end{align}
where $\{\mathbf{x}\}$ indicates the set of  elements in the vector
argument $\mathbf{x}$. As illustrated in Fig.\ref{fig1}, the min-sum
rule \eqref{minsum} disjointly exchanges the messages between
onefactor layer and two different variable layers where the factor
nodes are  effective only with the index $(k,l) \in \Omega$, and the
variable nodes correspond to the index $(i,j) \in \Lambda$.  Then,
totally, $4MN$ messages are handled per iteration since there are
$M$ effective factors, each of which generates $2N$ messages, and
$2N$ variables, each of which produces $M$ messages.


At the fixed-point, the scalar MAP estimate of each signal is
approximated by
\begin{align}\label{MAPest}
&\widehat x_{a,ji} = \arg \mathop {\min }\limits_{{x_{a,ji}} \in
\mathbb{R}} {\alpha _{ji}}({x_{a,ji}}),\,\,\,\widehat x_{b,ji} =
\arg \mathop {\min }\limits_{x_{b,ji} \in \mathbb{R}} {\beta
_{ji}}({x_{b,ji}}),
\end{align}
where the functions ${\alpha _{ji}}({x_{a,ji}}),{\beta
_{ji}}({x_{b,ji}})$, which include the posterior information, are
given by
\begin{align}
&\alpha_{ji }({x_{a,ji}}) = \log {f_{{{\mathbf{X}}_a}}}(\mathbf{x}_a
;{\mathcal{T}}_a) +
\sum\limits_{(k,l)\in \Omega} {{\mu_{kl \to ji}}({x_{a,ji}})},\label{posterior1}\\
&\beta_{ji }({x_{b,ji}}) \,= \log
{f_{{{\mathbf{X}}_b}}}(\mathbf{x}_b;{\mathcal{T}}_b) \,+
\sum\limits_{(k,l)\in \Omega} {{\nu_{kl \to ji
}}({x_{b,ji}})}.\label{posterior2}
\end{align}
In the AMP literature \cite{AMP}-\cite{Montanari}, the second term
of \eqref{posterior1},\eqref{posterior2} are handled as Gaussian
exponents. This is based on the assumption that the factor graph connection, by
the matrix $\mathbf{A}$, is sufficiently dense such that the
messages ${\mu_{kl \to ji}}({x_{a,ji}}),{\nu_{kl \to ji}}({x_{b,ji}})$
from each factor are Gaussian distributed by the law of large
numbers.

\begin{algorithm}[!t]
 \footnotesize
\caption{MixAMP Method }\label{algo1}
\begin{algorithmic}[0]
\Require Measurement matrix $\mathbf{A} \in \mathbb{R}^{ N \times
N}$,   Measurements $\mathbf{Y} \in \mathbb{R}^{ N \times N}$,
Undersampling operator $\mathcal{P}_{\Omega}\{\cdot\}$

\Ensure Recovered signals $\mathbf{X}_a^{t=t^*},\mathbf{X}_b^{t=t^*}
\in \mathbb{R}^{ N \times N}$
\\
\State {\bf{Initialization:}} ${{\mathbf{R}}^{t=0}} = {\mathbf{Y}}$,
${\theta ^{t=0}} = \frac{1}{M}\left\| {{{\mathbf{Y}}}} \right\|_F^2$
\State $ \,\,\,\,\,\,\,\,\,\,\,\,\,\,\,\,\,\,\,\,\,\,
\,\,\,\,\,\,\,\,\,\,\,\,\,\,\mathbf{X}_a^{t=0}=\mathbf{0},\mathbf{X}_b^{t=0}=\mathbf{0}$
\\
\For{$t=1$ \textbf{to} $t^*$ do}

\State ${\mathbf{X}}_a^t = \eta_a({{\mathbf{A}}^T}{{\mathbf{R}}^{t -
1}} {\mathbf{A}} + {\mathbf{X}}_a^{t - 1}; \theta^{t-1})$

\State ${\mathbf{X}}_b^t = {\eta
_b}({{\mathbf{A}}^T}{{\mathbf{R}}^{t - 1}}{\mathbf{A}} +
{\mathbf{X}}_b^{t - 1};{\theta ^{t - 1}})$

\State $\begin{gathered}
  {{\mathbf{R}}^t} = {\mathbf{Y}} - {\mathcal{P}_\Omega }\{ {\mathbf{A}}({\mathbf{X}}_a^t + {\mathbf{X}}_b^t){{\mathbf{A}}^T}\}  \hfill \\
  \,\,\,\,\,\,\,\,\,\,\,\,\,\,\,\,\,\,\,\,\,\, + {{\mathbf{R}}^{t - 1}}\begin{array}{l}\frac{N}{M}\end{array}\left\langle {{\eta _a}'({{\mathbf{A}}^T}{{\mathbf{R}}^{t - 1}}{\mathbf{A}} + {\mathbf{X}}_a^{t - 1};{\theta ^{t - 1}})} \right\rangle  \hfill \\
  \,\,\,\,\,\,\,\,\,\,\,\,\,\,\,\,\,\,\,\,\,\, + {{\mathbf{R}}^{t - 1}}\begin{array}{l}\frac{N}{M}\end{array}\left\langle {{\eta _b}'({{\mathbf{A}}^T}{{\mathbf{R}}^{t - 1}}{\mathbf{A}} + {\mathbf{X}}_b^{t - 1};{\theta ^{t - 1}})} \right\rangle  \hfill \\
\end{gathered}$

\State ${\theta ^t} = \frac{1}{M}\left\| {{{\mathbf{R}}^t}}
\right\|_F^2$ \EndFor
\end{algorithmic}
\end{algorithm}

The remaining steps for the AMP development consist of
\begin{enumerate}
\item \emph{the quadratic approximation step}, which approximates the min-sum
equations by quadratic functions, then converting \eqref{minsum} to
a parameter-passing rule whose messages are simple real numbers
(instead of functions),
\item \emph{the first-order
approximation step}, which cancels interference caused by the loopy
graph connection, and reduces the number of messages from $4MN$ to
$M+2N$.
\end{enumerate}
These development steps are conventional for the AMP methods applied
to the other types of linear inverse problems
\cite{AMP}-\cite{Montanari}, which are well formulated in the
literature \cite{Montanari}. Therefore, we omit details of such
remaining steps in this letter, immediately providing a final form
of  MixAMP in Algorithm \ref{algo1}.

One thing noteworthy  is that in Algorithm \ref{algo1} the two
disjoint iterations, described in Fig.\ref{fig1},  share the residual term $\mathbf{R}^t \in
\mathbb{R}^{ N \times N}$, which significantly reduces the number of the messages in the iteration. This can be explained using the two arguments given in \cite{Montanari}: \emph{i)} The messages,
sent by the $(k,l)$-th factor, have a residual form through the quadratic approximation step; for example, the message toward the variable index $(a,ji)$ is expressed as
\begin{align}\label{eq9}
r_{a, kl\to ji}=y_{kl}-\mathcal{P}_{\Omega}\left\{ \right.&\sum\nolimits_{j'i'\ne ji} a_{i'j'}a_{j'i'}x_{a,j'i' \to kl} \nonumber\\ &+\sum\nolimits_{ ji\in \Lambda} a_{ij}a_{ji}x_{b,ji \to kl}\left\}\right..
\end{align}
\emph{ii)} We can drop the directional dependency upon the destination index $(a,ji)$ in \eqref{eq9} by decomposing the residual message
$r_{a, kl\to ji}$ into a form of ``pure residual + directional correction" under the large limit $M,N \to \infty$,  \emph{i.e.},
\begin{align}\label{eq10}
r_{a, kl\to ji}&=r_{kl}+ \delta r_{a, kl\to ji}\nonumber\\
&=\underbrace{y_{kl}-\mathcal{P}_{\Omega}\{\sum\nolimits_{ji\in\Lambda} a_{ij}a_{ji}(x_{a,ji \to kl}+x_{b,ji \to kl})\}}_{=r_{kl}} \nonumber\\
&\,\,\,\,\,\,\,\,\,\,\,\,\,\,\,\,+ \underbrace{a_{ij}a_{ji}x_{a,ji}}_{=\delta r_{a, kl\to ji}},
\end{align}
where the correction term  $\delta r_{a, kl\to ji}$  has the order
of  $\mathcal{O}(N^{-1/2})$. The above two arguments equivalently
hold for the factor messages toward the index $(b,ji)$. Therefore,
the pure residual $r_{kl}^t \in \mathbf{R}^t$ is independent of the
index of the destination variable, enabling the residual sharing in
the two disjoint iteration.

The MixAMP  incorporates two distinct denoisers, denoted by
$\eta_a(\cdot),\eta_b(\cdot)$, according to sparsity types in the
mixture $\mathbf{X}_a + \mathbf{X}_b$. These denoisers undertake the
sub-optimization tasks given in \eqref{MAPest}, generating the MAP
estimate $\mathbf{X}_a^t,\mathbf{X}_b^t$ at every iteration. Hence,
choice of the sparsifying priors,
$f_{\mathbf{X}_a}(\cdot),f_{\mathbf{X}_b}(\cdot)$, determines
functional form of the denoisers; for some choices, we may need to
utilize external numerical solvers for the denoiser implementation
due to analytical difficulties of the prior exponent \emph{e.g.},
non-scalability and non-smoothness. This denoising concept of MixAMP
is analogous to the shrinkage in the context of \emph{iterative
shrinkage-thresholding} (IST). However, they are different in that
the MixAMP  denoisers shrink $\mathbf{X}_a^t,\mathbf{X}_b^t$ in the
domain of $\mathcal{T}_a$ and $\mathcal{T}_b$ respectively, whereas
the shrinkage operator of IST takes soft-thresholding in the
standard domain.


\begin{figure}
\centering
\includegraphics[width=8.6cm]{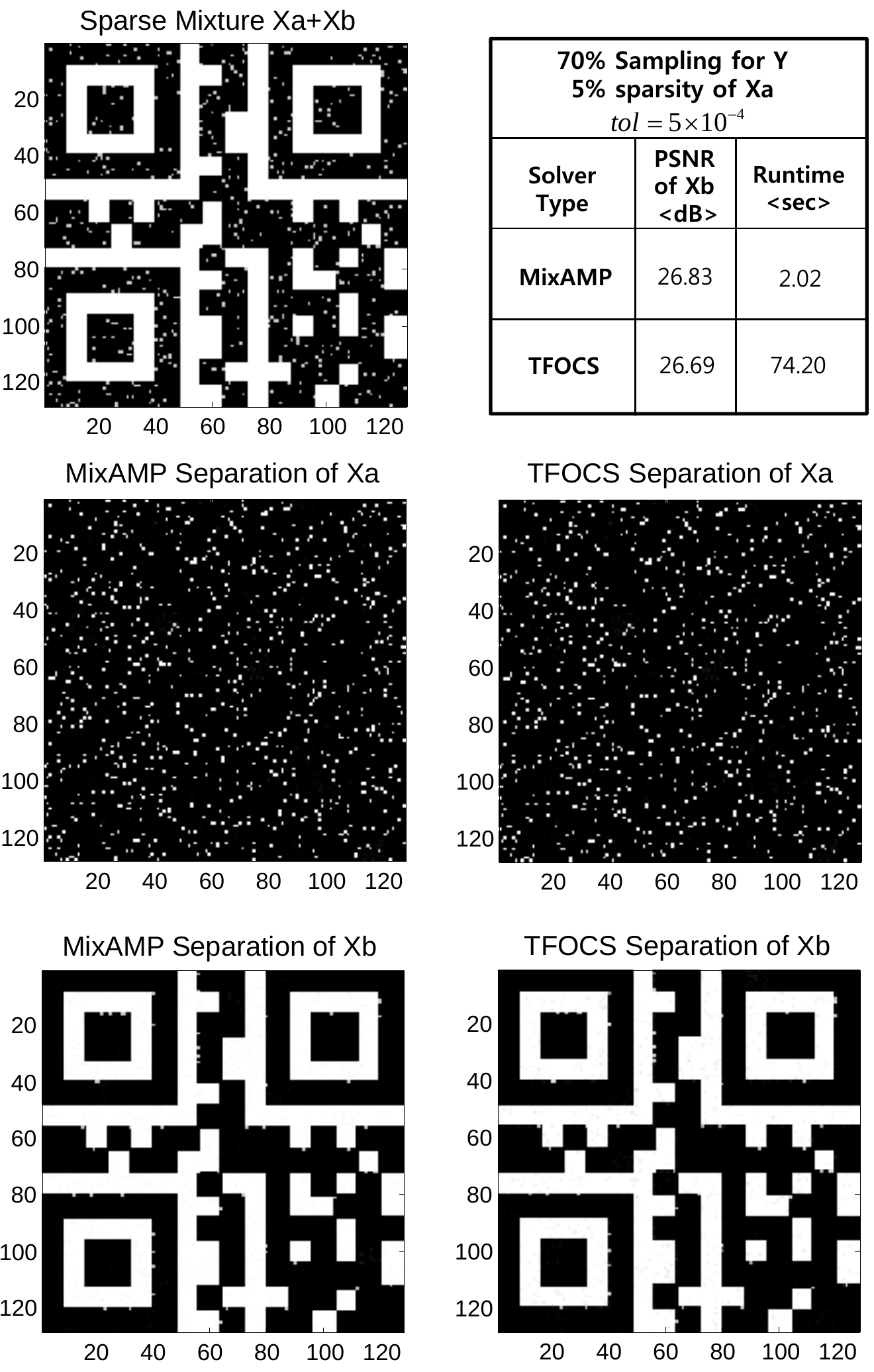}
\caption{An exemplary comparison of MixAMP and TFOCS in the
direct-and-group sparsity separation where we reconstruct a $128
\times 128$ QRcode image $\mathbf{X}_b$ by removing  shot noise
$\mathbf{X}_a$ (5\% sparsity)  given the measurements $\mathbf{Y}$
with $70\%$ sampling $(M/N=0.7)$. Here, we set
$\varepsilon=10^{-10},\lambda_1=0.5,\lambda_2=1.2$ for the TFOCS
method \eqref{TFOCS1}.} \label{fig2}
\end{figure}

\section{Case Studies for 2D Sparsity Separation}
This section provides an exemplary comparison of the MixAMP and the
TFOCS \cite{TFOCS} methods for  two different cases of the sparsity
separation. In each case, we first introduce denoisers applied to
the MixAMP method, and then we provide the separation example. We
inform that the TFOCS method is not directly applicable to the 2D
model \eqref{lin_sys}, which should be accompanied by the
vectorization of \eqref{lin_sys_vec}. This comparison is based on
the standard
 Gaussian  matrix $\mathbf{A}$ whose entries are \emph{i.i.d.} drawn from $\mathcal{N}(0,\frac{1}{M})$, and MATLAB
2014a with a 2.67-GHz Intel Quad Core \emph{i5} was used to generate
the results. In addition,   we stop the MixAMP and  TFOCS iteration
when $tol:=\frac{{\sqrt {||{\mathbf{X}}_a^{t - 1} -
{\mathbf{X}}_a^t||_F^2 + ||{\mathbf{X}}_b^{t - 1} -
{\mathbf{X}}_b^t||_F^2} }}{{\sqrt {||{\mathbf{X}}_a^t||_F^2 +
||{\mathbf{X}}_b^t||_F^2} }} \leq 5 \times 10^{-4}$ is met.

\subsection{Separation of Direct and Group Sparsity}
We consider  separation of a 2D direct-and-group sparse mixture. For
the direct sparsity denoiser, we apply the \emph{soft-thresholding}
which has been the most widely used because of  its simplicity
\cite{AMP}. Let $\mathbf{X}_a \in \mathbb{R}^{ \sqrt{N} \times
\sqrt{N} }$ have the direct sparsity. Then,  we can estimate
$X_{a,ji} \in \mathbb{R}$ via a scalable denoiser $\eta_a(\cdot):
\mathbb{R} \to \mathbb{R}$, given by
\begin{align}\label{st_denoiser}
\widehat x_{a,ji}={\eta _a}(x ;\theta )&:= \arg \mathop {\min }\limits_{{{X}_{a,ji}}} |{X_{a,ji}}| + \frac{\lambda }{2}({X_{a,ji}} - x )^2  \nonumber\\
&= \operatorname{sgn} (x)\max \{ |x| - \theta ,0\}.
\end{align}
In \eqref{st_denoiser}, the input $x \in \mathbb{R}$ is a Gaussian
variable corresponding to the second term of the posterior function
\eqref{posterior1}. This Gaussianity let the denoiser to solve a
penalized least squares problem, which is common  in the AMP
denoisers. In addition, it is well known that the soft-thresholding
is based on the Laplace prior; therefore, we have $\log
{f_{{{\mathbf{X}}_a}}}({{\mathbf{x}}_a}) = ||\mathbf{X}_a||_1$ in \eqref{posterior1}.

We adopt the \emph{block soft-thresholding} for the
group sparsity pursuit \cite{TVAMP}.  Let $\mathbf{X}_b \in
\mathbb{R}^{ \sqrt{N} \times \sqrt{N} }$ have the group sparsity.
Then, the block soft-thresholding denoiser, $\eta_b(\cdot):
\mathbb{R}^{ \sqrt{B} \times \sqrt{B} }\to \mathbb{R}^{ \sqrt{B}
\times \sqrt{B} }$, is given by
\begin{align}\label{block_th}
\widehat {\mathbf{x}}_{b,B}={\eta _b}({{\mathbf{x}}_B};\theta )&: = \arg \mathop {\min }\limits_{{{\mathbf{X}}_{b,B}}} ||{{\mathbf{X}}_{b,B}}||_F + \frac{\lambda }{2}||{{\mathbf{X}}_{b,B}} - {{\mathbf{x}}_B}||_F^2 \nonumber\\
&= {{\mathbf{x}}_B} \cdot \max \{ 1 - \theta
||{{\mathbf{x}}_B}||_F^{ - 1},0\}
\end{align}
where the input $\mathbf{x}_B \in \mathbb{R}^{ \sqrt{B} \times
\sqrt{B} }$ is a block matrix such that the signal $\mathbf{X}_b$ is
partitioned into $N/B$ square blocks with the size $B$.
Specifically, this block denoiser makes its block argument
$\mathbf{x}_B$ to a zero matrix if $||{{\mathbf{x}}_B}||_F \leq
\theta$, otherwise diminishing it by the quantity $\theta$ to  the
origin. In addition, this block thresholding is related to  a block
Gaussian prior; hence, we have $\log
{f_{{{\mathbf{X}}_b}}}({{\mathbf{x}}_b}) = \sum\nolimits_{{\text{All
blocks}}} {||{{\mathbf{X}}_{b,B}}||_F}$ from \eqref{posterior2}.

Fig.\ref{fig2} displays a separation example by MixAMP and TFOCS,
where we reconstruct a $128 \times 128$ QRcode image $\mathbf{X}_b$
by removing  shot noise $\mathbf{X}_a$  given the measurements
$\mathbf{Y}$ with $70\%$ sampling $(M/N=0.7)$. The QR code  refers
to the group sparsity part of the mixture image, while the shot
noise  models the direct sparsity part. In the TFOCS method, we
recast \eqref{L1min} to \emph{Combining-$L_1$-and-Group
minimization}, solving
\begin{align}\label{TFOCS1}
\begin{gathered}
  \min \,\,\,\lambda_1||{{\mathbf{X}}_a}|{|_1} + \lambda_2\sum\limits_{{\text{\tiny All blocks}}} {||{{\mathbf{X}}_{b,B}}|{|_F}}  \hfill \\
  {\text{s.t.}}\,\,\,\,\,||{\mathbf{Y}} - \mathcal{P}_{\Omega}\{ {\mathbf{A}}({{\mathbf{X}}_a} + {{\mathbf{X}}_b}){\mathbf{A}^T}\} ||_F^{} \leq \varepsilon  \hfill \\
\end{gathered}
\end{align}
where $\varepsilon,\lambda_1,\lambda_2\geq 0$ are calibration
scalars. The result of Fig.\ref{fig2} reports that MixAMP is much
faster than TFOCS in CPU runtime while providing comparable
reconstruction quality in PSNR.

\subsection{Separation of Direct and Finite-Difference Sparsity}
We present another case  by introducing finite-difference (FD)
sparsity, then addressing a separation problem with a direct-and-FD
sparse mixture. For the FD sparsity pursuit, we apply the
\emph{total variation} (TV) denoiser  \cite{TVAMP} which has been
investigated by the numerous literature (see for example \cite{TV}).
The TV denoiser is neither scalable nor block-separable because the
FD sparsity cannot be defined by a single scalar of the signal, but depending upon
all the adjacent of the scalar. Let $\mathbf{X}_b \in \mathbb{R}^{ \sqrt{N}
\times \sqrt{N} }$ have the FD sparsity. Then, we consider the TV
denoiser, $\eta_b(\cdot): \mathbb{R}^{ \sqrt{N} \times \sqrt{N} }\to
\mathbb{R}^{ \sqrt{N} \times \sqrt{N} }$, which solves
\begin{align}\label{TVdenoiser}
{\widehat {\mathbf{x}}_b} = {\eta _b}({\mathbf{x}};\lambda) : = \arg
\mathop {\min }\limits_{{{\mathbf{X}}_b}} ||
\mathbf{X}_b||_{\text{TV}}+ \frac{\lambda }{2}||{{\mathbf{X}}_b} -
{\mathbf{x}}||_F^2.
\end{align}
It is recognizable from  \eqref{posterior2} that the TV norm $||
\mathbf{X}_b||_{\text{TV}}$ is the exponent of the sparsifying prior
such that $\log {f_{{{\mathbf{X}}_b}}}({{\mathbf{x}}_b})
=||\mathbf{X}_b||_{\text{TV}}$. For the implementation of
\eqref{TVdenoiser}, external numerical solvers have been mostly
considered  since the TV norm is analytically non-scalable and
non-smooth \cite{TVAMP}.

\begin{figure}
\centering
\includegraphics[width=8.5cm]{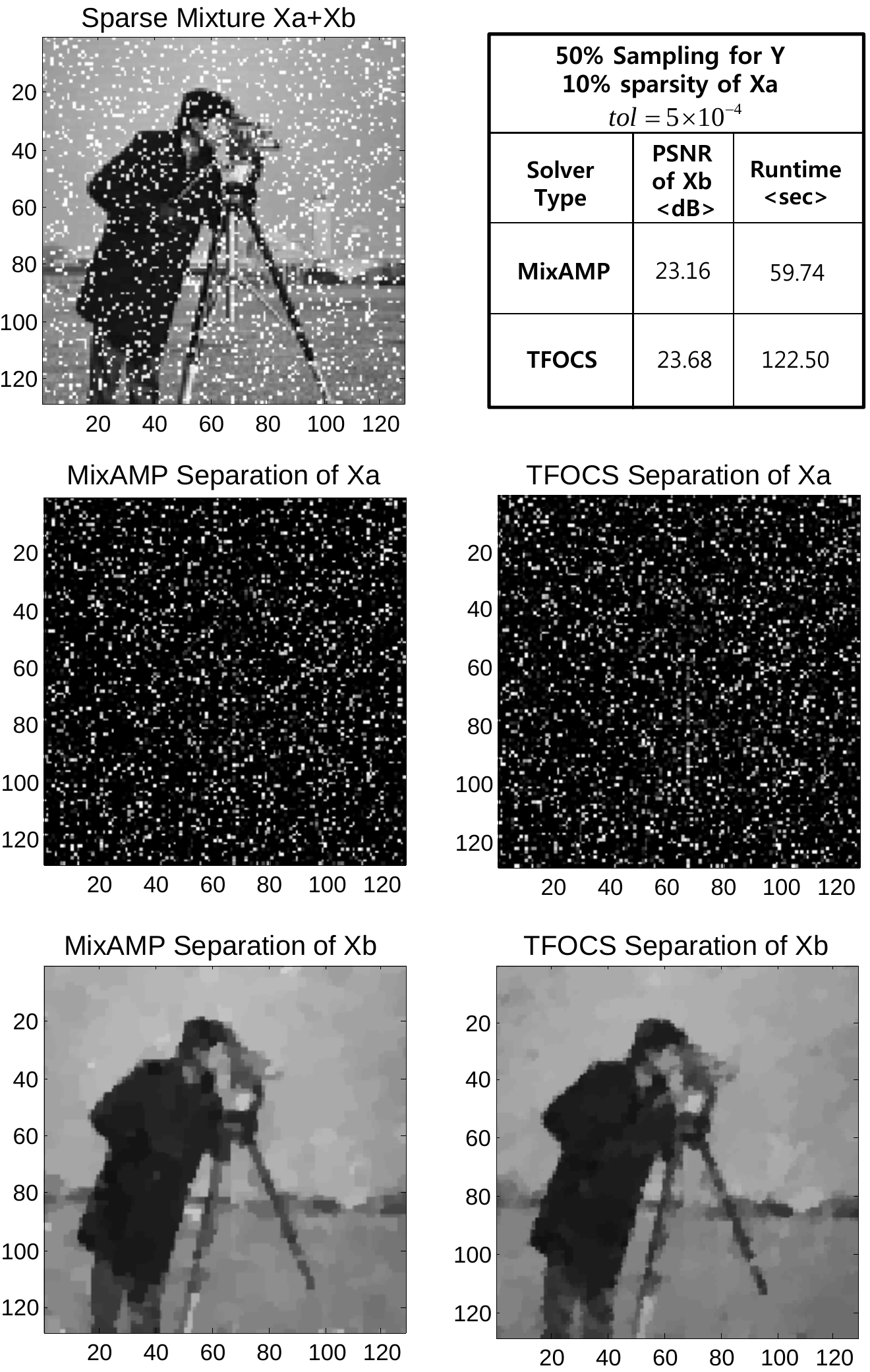}
\caption{An exemplary comparison of MixAMP and TFOCS in the
direct-and-FD sparsity separation where we simultaneously estimate a
shot noise $\mathbf{X}_a$ (10\% sparsity) and a $128 \times 128$
\emph{Cameraman} image $\mathbf{X}_b$ from the 2D measurements
$\mathbf{Y}$ with $50\%$ sampling $(M/N=0.5)$. Here, we set
$\varepsilon=10^{-10},\lambda_1=2.0,\lambda_2=1.4$ for the TFOCS
method \eqref{TFOCS2}.} \label{fig3}
\end{figure}

In Fig.\ref{fig3}, we simultaneously estimate a shot noise
$\mathbf{X}_a$ (10\% sparsity) and a $128 \times 128$
\emph{Cameraman} image $\mathbf{X}_b$ from the 2D measurements
$\mathbf{Y}$ with $50\%$ sampling $(M/N=0.5)$. In this example, we
postulate that the \emph{Cameraman} image has the FD sparsity and
the shot noise is directly sparse. For the MixAMP separation, we
apply the soft-thresholding \eqref{st_denoiser} for $\mathbf{X}_a$,
implementing the anisotropic TV denoising \eqref{TVdenoiser} for
$\mathbf{X}_b$ using the 2D-Bregman iteration (Section 4 of
\cite{bregman}). In the TFOCS method, we recast \eqref{L1min} to
\emph{Combining-$L_1$-and-TV minimization}, solving
\begin{align}\label{TFOCS2}
\begin{gathered}
  \min \,\,\lambda_1||{{\mathbf{X}}_a}|{|_1} + \lambda_2||{{\mathbf{X}}_b}|{|_{\text{TV}}} \ \hfill \\
  {\text{s.t.}}\,\,\,\,||{\mathbf{Y}} - \mathcal{P}_{\Omega}\{ {\mathbf{A}}({{\mathbf{X}}_a} + {{\mathbf{X}}_b}){\mathbf{A}^T}\} ||_F \leq \varepsilon.  \hfill \\
\end{gathered}
\end{align}
Likewise to the example in Section III-A, the result of
Fig.\ref{fig3} validates the computational advantages of MixAMP
where
 MixAMP is approximately 2 time faster
than  TFOCS  for the same task. The scale difference of the runtime
gap from the result of Section III-A is coming from the fact that
the Bregman-TV denoiser \eqref{TVdenoiser} requires more
computations than the block soft-thresholding \eqref{block_th} does.

These two comparison results support our claim that MixAMP outperforms the TFOCS method in computational efficiency
 under the 2D sparse signal separation task.

\section{Conclusions}
In this letter, we have discussed the MixAMP method for  the
ill-posed \emph{$L_1$-Split-Analysis}, applying the 2D sparse signal
separation problem. We first have developed the MixAMP method based
on the factor graphical model of the 2D CS measurement model
\eqref{lin_sys}. Then, we have provided two cases of the study for
the sparsity separation, validating the computational advantages of
MixAMP through exemplary comparisons to the conventional first-order
method, TFOCS \cite{TFOCS}. Therefore, we claim that MixAMP is a
very good alternative of the TFOCS method for solving the
\emph{$L_1$-Split-Analysis} in the 2D sparse separation problem.

%
%
%

\begin{figure*}[!t]
\centering
\includegraphics[width=18cm]{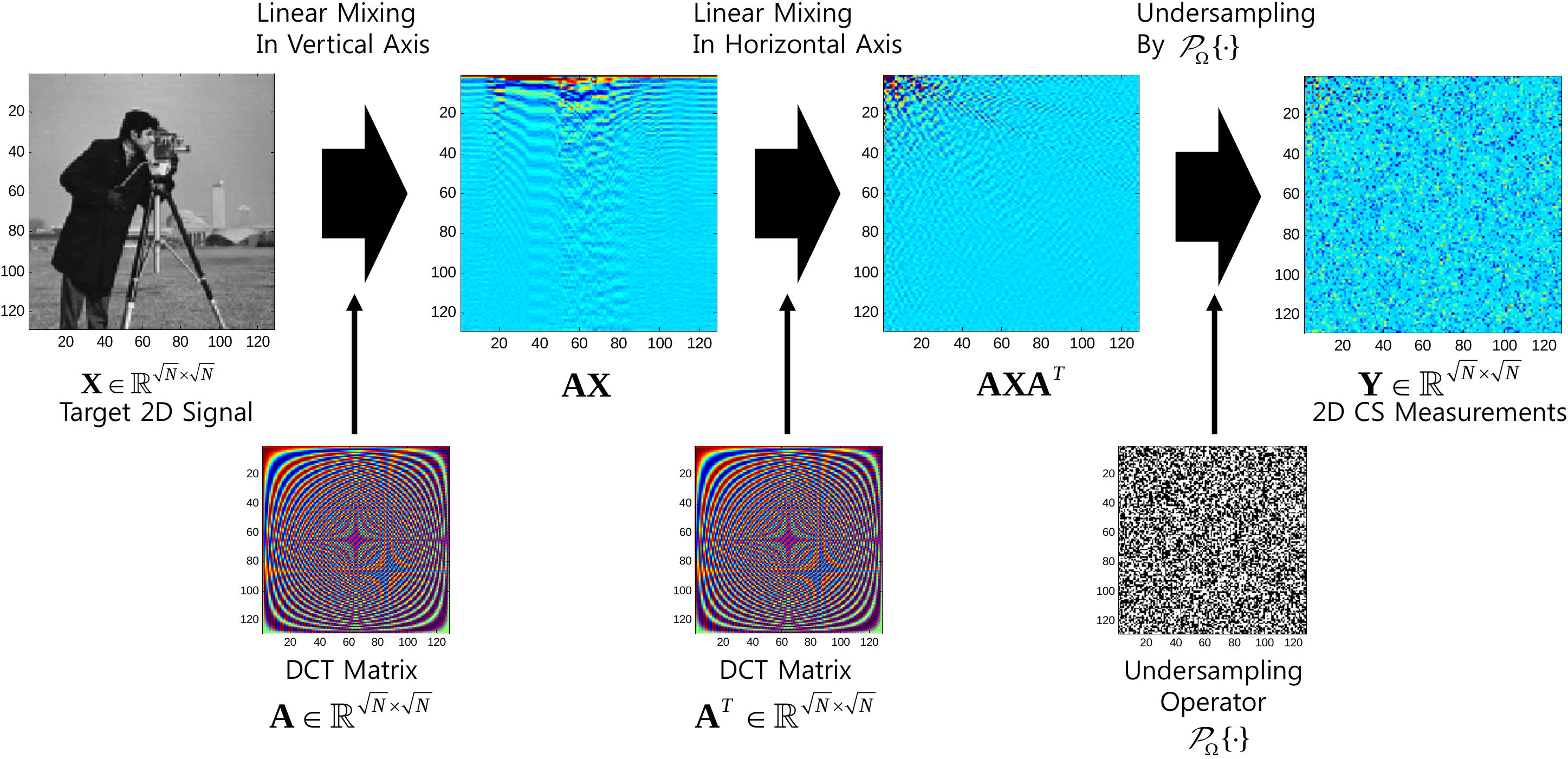}
\caption{An example of 2D CS measurement generation in the model
(1), where we consider the DCT matrix $\mathbf{A}$ and the $128
\times 128$ \emph{Cameraman} image as the 2D target signal
$\mathbf{X}$. First, the signal $\mathbf{X}$ is linearly mixed in
two perpendicular axes by the DCT matrix $\mathbf{A}$, generating
$\mathbf{AXA^T}$. Second, the measurement $\mathbf{Y}$ is produced
by applying the undersampling operator
$\mathcal{P}_{\Omega}\{\cdot\}$ to $\mathbf{AXA^T}$.}
\label{fig1_supp}
\end{figure*}

\section*{Supplementary Material}
\subsection{2D Compressed Sensing Model \eqref{lin_sys}: A Practical Viewpoint}

In this letter, we consider ``the compressed sensing (CS) technique"
with the model \eqref{lin_sys}. The model \eqref{lin_sys} requires
two-steps for the CS measurement generation of $\mathbf{Y}$ as shown
in Fig.\ref{fig1_supp}: 1) linear mixing in two perpendicular axes
by $\mathbf{AXA^T}$ and 2) undersampling by
$\mathcal{P}_{\Omega}\{\cdot\}$. One can argue that this two-steps
generation makes us to lose an advantage of CS which generates
$\mathbf{Y}$ in the first place. Nevertheless, we state that this
two-step generation is useful because of two practical reasons:
\begin{enumerate}
\item \emph{For fast measurement generation with unitary matrices $\mathbf{A}$}: The
model \eqref{lin_sys} can provide an accelerated generation of
$\mathbf{Y}$ when the matrix $\mathbf{A}$ is some unitary types. For
instance, this acceleration  utilizes \emph{Fast Cosine Transform}
(FCT) when $\mathbf{A}$ is the DCT matrix, and \emph{Fast Fourier
Transform} when $\mathbf{A}$ is the Fourier matrix. Here, we
consider the DCT case as an example. Let
$\mathbf{A}_{\text{DCT}}\in\mathbb{R}^{\sqrt{N}\times\sqrt{N}}$
denote the DCT matrix and
$\mathbf{A}_{\text{subDCT}}\in\mathbb{R}^{\sqrt{M}\times\sqrt{M}}$
is a sub-DCT matrix whose rows are randomly sampled from
$\mathbf{A}_{\text{DCT}}$. The conventional CS generation, expressed
as
\begin{align}\label{convCS}
\mathbf{Y}=\mathbf{A}_{\text{subDCT}}\,\mathbf{X}\,\mathbf{A}_{\text{subDCT}}^T
\in\mathbb{R}^{\sqrt{M}\times\sqrt{M}},
\end{align}
spends $\mathcal{O}(NM^{\frac{1}{2}})$ computations due to the
matrix multiplications. Compared to \eqref{convCS}, the FCT-based
generation with the model \eqref{lin_sys}, \emph{i.e.},
\begin{align}
\mathbf{Y}&=\mathcal{P}_{\Omega}\{\mathbf{A}_{\text{DCT}}\,\mathbf{X}\,\mathbf{A}_{\text{DCT}}^T
\},\nonumber\\
&=\mathcal{P}_{\Omega}\{\text{FCT}_{2D}[\mathbf{X}]\}\in\mathbb{R}^{\sqrt{N}\times\sqrt{N}},
\end{align}
is  computationally efficient with $\mathcal{O}(N \log
N^{\frac{1}{2}})$; therefore, its efficiency gets remarkable as the
system size $N$ increases. In addition, it is obvious that the FCT
method is not applicable with the sub-DCT matrix
$\mathbf{A}_{\text{subDCT}}$.
\item   \emph{Use of a random decimation operator instead of $\mathcal{P}_{\Omega}\{\cdot\}$}:
In practice, the random undersampling operator
$\mathcal{P}_{\Omega}\{\cdot\}$  can be simply replaced by a random
decimation operator. Then, the  samples of $\mathbf{Y}$ can be
stored to a memory with the size $M$ by holding the knowledge of the
set $\Omega$. In this case, however, we need an inverse operator of
the decimation operator to calculate
$\mathbf{A}^T\mathbf{R}\mathbf{A}$ for the MixAMP method.
\end{enumerate}

\subsection{Hardship of Sparse Separation Problem}
The sparsity separation, handled in this letter, is basically harder
problem than the conventional single sparsity recovery. This is
because the sparsity  separation  includes not only the
reconstruction of the mixture signal $\mathbf{X}$ from the CS
measurements $\mathbf{Y}$, but also the separation of the two
sparsity, $\mathbf{X}_a$ and $\mathbf{X}_b$, from the mixture
$\mathbf{X}$. We can look at this hardship of the sparsity
separation by simple
 manipulation from \eqref{lin_sys}:
\begin{align}\label{mani}
{\mathbf{Y}} = {\mathcal{P}_\Omega }\left\{
{[{\mathbf{A}}|{\mathbf{A}}]\left[
\begin{gathered}
  {{\mathbf{X}}_a} \hfill \\
  {{\mathbf{X}}_b} \hfill \\
\end{gathered}  \right]{{\mathbf{A}}^T}} \right\},
\end{align}
From \eqref{mani}, we can reasonably conjecture that the
reconstruction of the concatenated signal $[\mathbf{X}_a |
\mathbf{X}_b]^T$ from ${\mathbf{Y}}$ requires larger $M$ than the
single sparsity recovery does. In addition, while revising this
letter, we became aware of a theoretical work of Studer \emph{et
al.} \cite{studer1} which supports our conjecture by providing a
coherence-based sufficient condition for recovery guarantees of
${\mathbf{X}}_a$ and ${\mathbf{X}}_b$.

\begin{figure*}[!t]
\centering
\includegraphics[width=18cm]{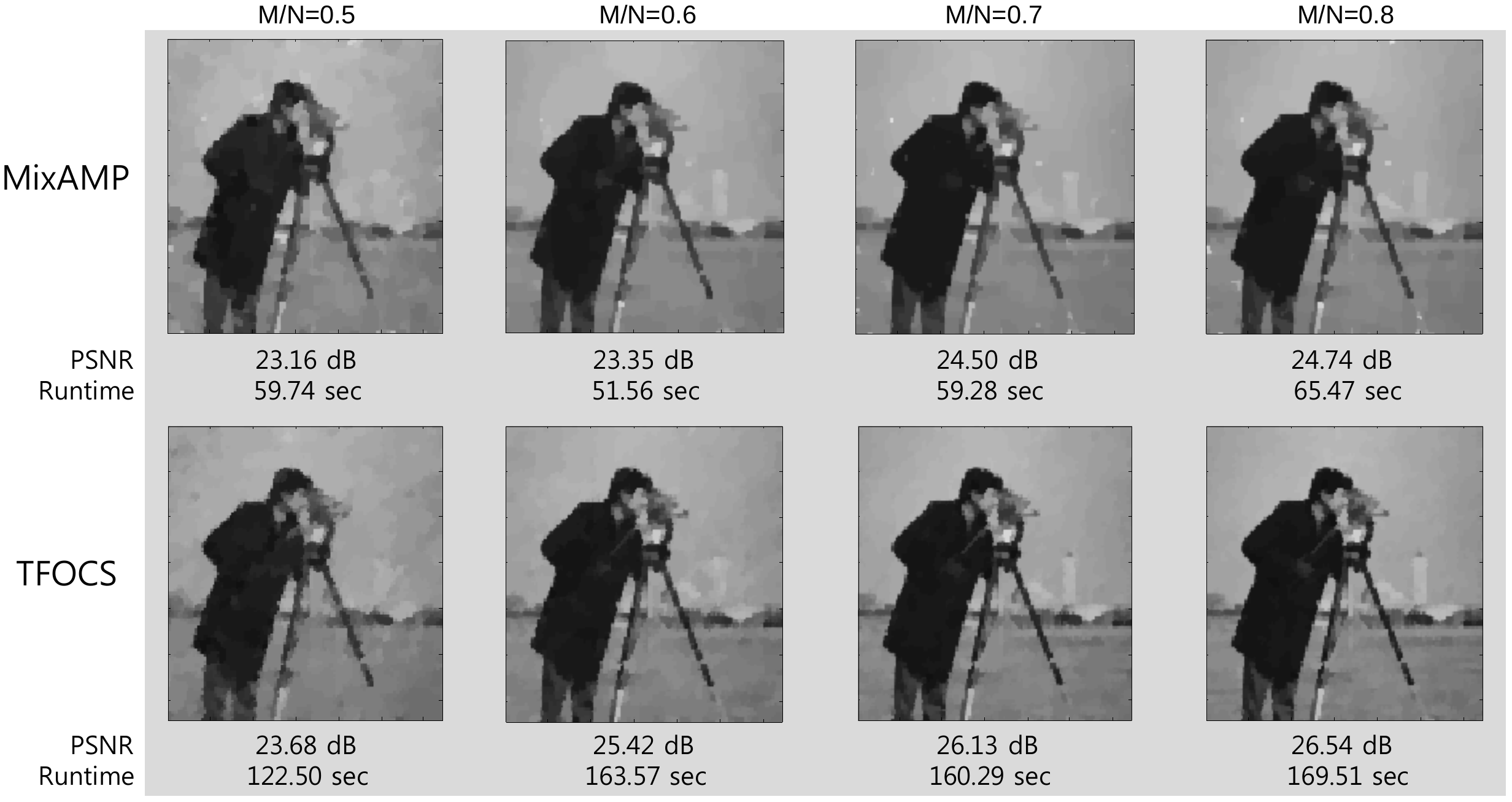}
\caption{Extended numerical result of the directed-and-FD sparsity
separation in Section III-B: In this figure, we only plot
reconstruction of the $128 \times 128$ \emph{Cameraman image}
$\mathbf{X}_b$ for a variety of the sampling rate $M/N$. The others
for the experimental setup remains the same as Section III-B. For
the TFOCS method (15), we use empirically-tuned parameter set
$(\lambda_1,\lambda_2,\varepsilon)$.} \label{fig2_supp}
\end{figure*}

Naturally, the reconstruction quality of the sparse separation gets
better as the sampling rate $M/N$ increases. In Fig.\ref{fig2_supp},
we provide an extended result of the $\mathbf{X}_b$ reconstruction
in the directed-and-FD sparsity separation (Section III-B) for a
variety of $M/N$. From the figure, we see that the both methods
improve their reconstruction quality given a higher $M/N$. This
result also shows that although the TFOCS method has a small lead in
the reconstruction quality as $M/N$ becomes higher, the MixAMP
method is still remarkably advantageous in computational cost.

\end{document}